\documentstyle[11pt,paspconf,psfig]{article}

\begin{document}

\title{Completeness and confusion in the identification 
       of Lyman-break galaxies}

\author{Garret Cotter$^{\! 1}$, Toby Haynes$^{\! 1}$, Joanne
C. Baker$^{\! 1, 2}$, \\
Michael E. Jones$^{\! 1}$, Richard Saunders$^{\! 1}$}

\affil{1 Astrophysics, Cavendish Laboratory,\\
Madingley Road, Cambridge, CB3 0HE, UK \\
2 Astronomy Department, University of California,\\ 
Berkeley CA 94720, USA}

\begin{abstract}

We have carried out a study to simulate distant clusters of galaxies
in deep ground-based optical images. We find that when model galaxies
are added to deep images obtained with the William Herschel Telescope,
there is considerable scatter of the recovered galaxy colours away
from the model values; this scatter is larger than that expected from
photometric errors and is significantly affected by confusion, due to
ground-based seeing, between objects in the field. In typical
conditions of $\approx$ 1-arcsec seeing, the combination of confusion
and incompleteness causes a considerable underestimation of the true
surface density of $z~\approx~3$ galaxies. We argue that the actual
surface density of $z~\approx~3$ galaxies may be several times greater
than that estimated by previous ground-based studies, consistent with
the surface density of such objects found in the HDF.

\end{abstract}


\section{Introduction}

The field of the of the $z = 3.8$ quasar pair PC1643+4631 A \& B
contains a Cosmic Microwave Background decrement (Jones et al. 1997)
which may be the Sunyaev-Zel'dovich effect of a cluster of galaxies at
$ z >> 1$ (Saunders et al. 1997; Kneissl et al. 1998).  In an attempt
to detect such a cluster, we have carried out deep $UGVRI$ imaging of
the field. No cluster is immediately obvious in the images, so we
carried out Monte-Carlo simulations to quantify our ability to detect
a cluster of galaxies in our images (full details are given in Haynes
et al. 1999 and Cotter et al. 1999).

\section{Model high-$z$ cluster galaxies}

Model clusters were created using simulated colours of evolving
galaxies in the redshift range $0 < z < 4$, and added to our WHT
images. We then used photometric redshift techniques to try to recover
the simulated cluster. As the cluster redshift reached $z \approx 1$
and beyond, the lack of strong spectral features in the optical made
the cluster increasingly difficult to detect. However, even at $z \sim
3$, where the characteristic Lyman-limit break became detectable, a
large fraction of the the fake cluster galaxies were still recovered
from the simulation with ambiguous colours (Fig 1). Indeed, in our $z
= 3.0$ simulation, only one in five of the model cluster galaxies was
identified as such by $UGR$ selection. The recovered colours were
skewed towards the red in $G - R$; this is a result of confusion with
other objects in the field.

\begin{figure}
\centerline{\psfig{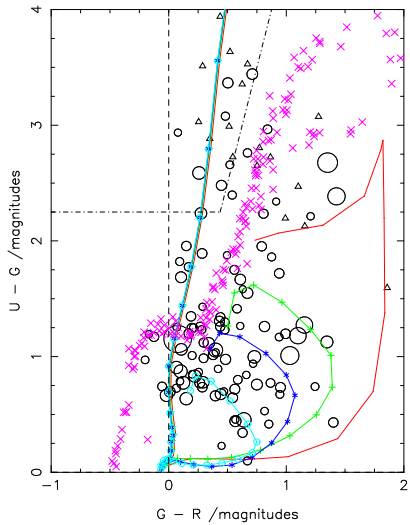}}
\caption{\small Recovered $UGR$ colours of simulated $2.5 < z < 3.5 $
cluster galaxies. All objects shown have measured $R < 25.5$ and $G >
2 \sigma$. The triangles denote 1-$\sigma$ lower limits in $U -
G$. The crosses are stars from the Gunn \& Stryker (1983) database and
the dot-dash line is our bound for selecting $z > 3$ candidates. The
tracks show the simulated galaxy colours at various redshifts; the
point at $U - G = 2.2$ and $G - R = 0.3$ corresponds to $z = 3.0$, and
points at larger $U - G $ are at higher redshift in steps of $\Delta z
= 0.1$. Note that the recovered colours are skewed towards the red in $G - R$.}
\end{figure}

\section{Recovery of model $ z \approx 3$ galaxies}

Our original search for $z \approx 3$ Lyman-break galaxies (LBGs) in
these images (Cotter \& Haynes 1998) had recovered a reasonable number
of candidates---approximately 1.1 arcmin$^{-2}$, similar to the
findings of the surveys of Steidel et al. (1996,1998). The fact that
our recovery of simulated high-$z$ cluster galaxies was so inefficient
therefore prompted us to measure the effects of completeness and
confusion specifically for $z \approx 3$ galaxies. 

We ran 1000 simulations, each time adding ten fake LBGs to our
images. Fake LBGs were drawn from a Schecter luminosity function with
$R_* = 24$ and $\alpha = 1.06$; all had input colours $G - R = 2.2$,
$U - G = 0.3$.  We used input half-light radii of 0.2-0.3$''$
(Giavalisco et al. 1998).

Then, using FOCAS, we attempted to recover the LBGs. Our selection
criteria are chosen to be as close as possible to that of Steidel et
al. We select only those galaxies clearly detected with $R < 25.5$
above the 3-$\sigma$ isophote in $R$, measure magnitudes in $U$ and
$G$ through this $R$-band isophote, and then impose a colour cut of $
U - G > 2$ , $U - G > 4(G - R) + 0.5$, which is closely equivalent to
the ``robust'' colour selection of Steidel et al. First, we find that
53\% of galaxies with input $R < 25.5$ are selected to the isophotal
$R = 25.5$ limit. Second, we find that, as for our fake cluster
galaxies, a large fraction of the fake LBGs are scattered far away
from their input colours.

In total, only 23\% of the input LBGs with $R < 25.5$ remain within
the $z \ge 3$ region of the $UGR$ plane (Fig 2). Therefore, the true
number of LBG candidates in our images may be four times greater than
the 1.1 arcmin$^{-2}$ we measure. Again we stress that, while our
$UGR$ filter set is slightly different from the $U_n G \mathcal{R}$
used by Steidel et al., our search for genuine LBG candidates in these
images finds a surface density at least as great as that of Steidel et
al.

\begin{figure}[ht]
\centerline{\psfig{file=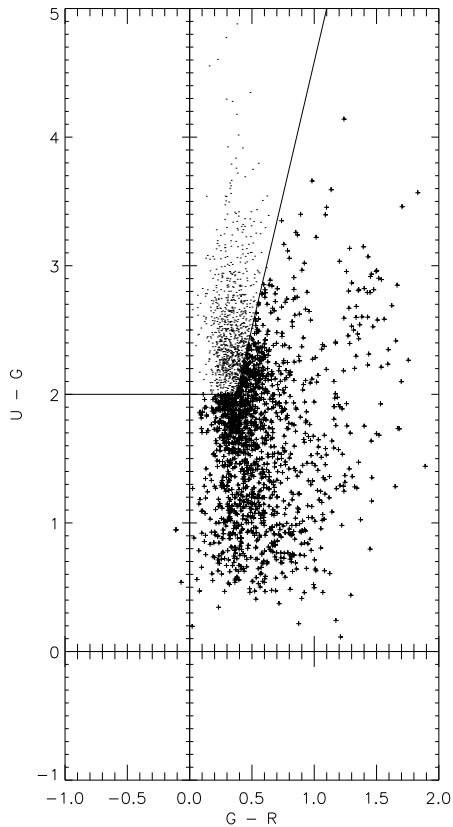,width=0.5\linewidth,clip=}}
\caption{\small Recovered colours of fake $z = 3.0$ LBGs; all have
input colours $G - R = 2.2$, $U - G = 0.3$. All galaxies recovered to
the $R=25.5$ 3-$\sigma$ isophote are shown; crosses mark those outside
the $z \ge 3$ region of colour-colour space defined by the solid line
($ U - G > 2, U - G > 4(G - R) + 0.5$). Dots mark the 23\% of the
galaxies with input $R < 25.5$ remaining in the $z \ge 3$ region.}
\end{figure}

These results suggest that ground-based $UGR$ selection, while
extremely successful at identifying $z \sim 3$ galaxies, may miss a
significant fraction of the population. This may have a bearing on the
apparent discrepancy between the surface densities of LBGs measured in
the Hubble Deep Field (HDF) and in ground-based surveys. There are 12
galaxies in the HDF which have spectroscopic redshifts $2.8<z<3.5$ and
$V_{606}<25.5$ (Dickinson 1997). These galaxies correspond to those
which would be detected by the ``robust'' LBG candidate criteria of
Steidel et al. (1998). However, one would expect, given the published
surface-density of ``robust'' LBG candidates of $\approx0.7$
arcmin$^{-2}$ (Steidel et al. 1998), to find only three such LBGs in
the HDF. Of course, cosmic variance will be significant for the HDF;
but it is striking that the apparent overdensity of LBGs in the HDF
corresponds with our estimate of the fraction of LBGs lost to
incompleteness and confusion in the ground-based images.

\section{Conclusions}

We have carried out simulations to examine the effectiveness of
searches for high-redshift galaxies in deep optical images typical of
those obtained with 4-m telescopes. We find that the scatter of the
recovered colours of model galaxies away from their model colours is
two to three times greater than that expected from the photometric
errors alone and arises as a result of confusion between the simulated
galaxies and the real objects in the field. Because of the effects of
incompleteness and confusion, the surface densities of LBGs based on
ground-based imaging may be underestimated by a factor of four; this
is consistent with the surface density of LBGs measured in the HDF.

To investigate these effects further, we will carry out more detailed
simulations using our present data to investigate how the inferred
luminosity function of the LBGs is affected by incompleteness and
confusion. We also plan deep imaging programmes with the new
generation of wider-field, high-image-quality instrumentation as it
becomes available.

\end{document}